# COVRECON: Combining Genome-scale Metabolic Network Reconstruction and Data-driven Inverse Modeling to Reveal Changes in Metabolic Interaction Networks

Jiahang Li[1], Steffen Waldherr[1], Wolfram Weckwerth[1,2,*]

[1]Molecular Systems Biology Lab (MOSYS), Department of Functional and Evolutionary Ecology, University of Vienna, Djerassiplatz 1, 1030 Vienna, Austria

[2]Vienna Metabolomics Center (VIME), University of Vienna, Djerassiplatz 1, 1030 Vienna, Austria

*To whom correspondence should be addressed.

**Contact:** Jiahang Li (jiahangl93@univie.ac.at), Steffen Waldherr (steffen.waldherr@univie.ac.at) and Wolfram Weckwerth (wolfram.weckwerth@univie.ac.at)

## Abstract

One central goal of systems biology is to infer biochemical regulations from large-scale OMICS data. Many aspects of cellular physiology and organism phenotypes could be understood as a result of the metabolic interaction network dynamics. Previously, we have derived a mathematical method addressing this problem using metabolomics data for the inverse calculation of a biochemical Jacobian network. However, these algorithms for this inference are limited by two issues: they rely on structural network information that needs to be assembled manually, and they are numerically unstable due to ill-conditioned regression problems, which makes them inadequate for dealing with large-scale metabolic networks.

In this work, we present a novel regression-loss based inverse Jacobian algorithm and related workflow COVRECON. It consists of two parts: a, Sim-Network and b, Inverse differential Jacobian evaluation. Sim-Network automatically generates an organism-specific enzyme and reaction dataset from Bigg and KEGG databases, which is then used to reconstruct the Jacobian's structure for a specific metabolomics dataset. Instead of directly solving a regression problem, the new inverse differential Jacobian part is based on a more robust approach and rates the biochemical interactions according to their relevance from large-scale metabolomics data.

This approach is illustrated by in silico stochastic analysis with different-sized metabolic networks from the BioModels database. The advantages of COVRECON are that 1) it automatically reconstructs a data-driven superpathway metabolic interaction model; 2) more general network structures can be considered; 3) the new inverse algorithms improve stability, decrease computation time, and extend to large-scale models.

**Graphic Abstract**

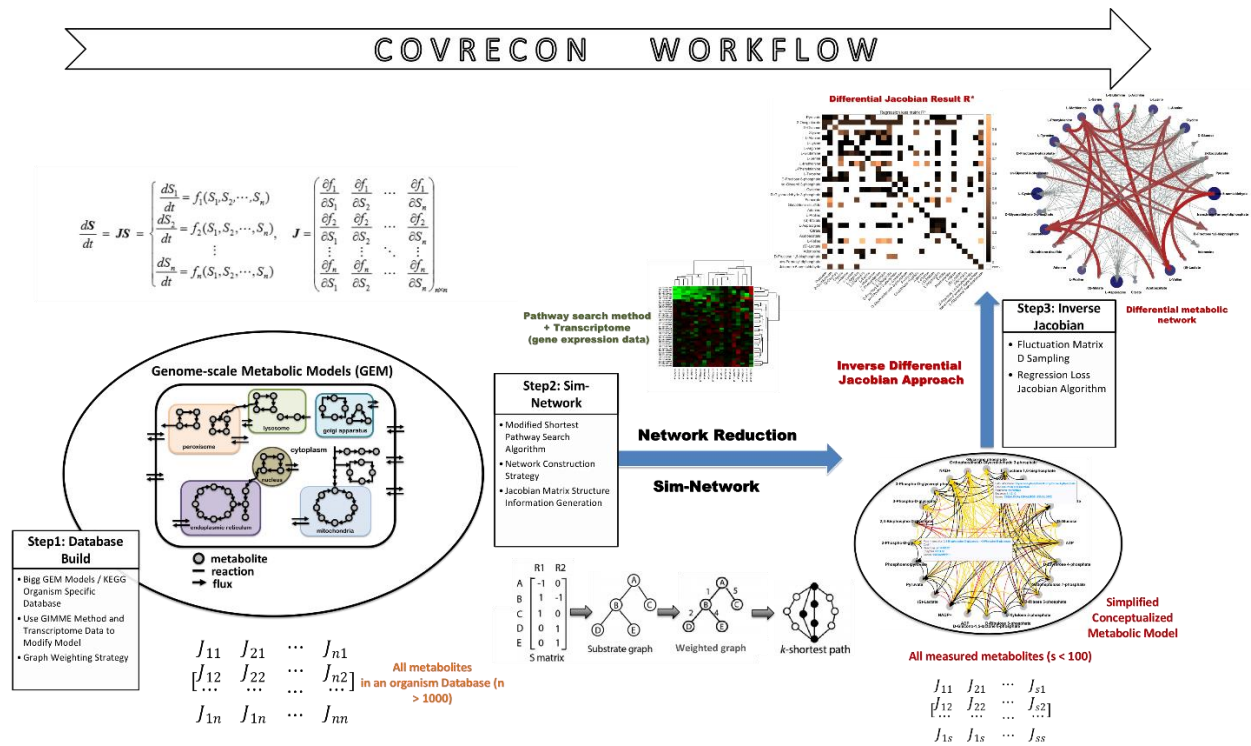

# 1 Introduction

Recent studies in system biology generate large datasets of molecular, genomic, and physiological variables, with the aim to understand complex diseases and regulatory interactions in biochemical networks from clinical studies (Elgendy, et al., 2019; Linke, et al., 2017). However, the functional interpretation of such datasets and the inference of how regulatory mechanisms in the underlying biochemical networks change in disease conditions relies on developing proper mathematical analysis and inference methods (Weckwerth, 2010; Weckwerth, 2011; Weckwerth, 2019).

The primary method in genome-scale metabolomics data processing is statistics. In recent studies, both conventional and machine learning methods have been implemented into metabolomics data processing. The conventional statistical methods such as t-test, clustering, and principal component analysis (PCA) are widely used but are not able to reveal the underlying biochemical regulatory interactions. Nevertheless, many recently developed statistics and computer science techniques have primarily enhanced the statistical power in metabolomics data analysis, such as deep learning (LeCun, et al., 2015), genetic algorithms (Mitchell, 1998), and boosting machine learning methods (Chen and Guestrin, 2016). Yet these methods provide limited insight into how the information in a biochemical network is transferred, what the critical regulatory steps are, and how a regulatory mechanism changes under different conditions. Here, a metabolic interaction network is defined from biochemical or regulatory interactions between identified metabolites. Interactions can be direct or involve several connected reactions (superpathways) (Nägele, et al., 2014; Weckwerth, 2019). As shown in figure 1a, many aspects of cellular physiology and organism phenotype could be understood as a result of the metabolic interaction network dynamics (De Martino, et al., 2018). Thus analyzing the changes in a metabolic interaction network for different phenotypes can give insight into changes of the underlying regulatory mechanisms. This interaction difference is influenced by complex aspects, where reaction enzyme activity difference plays a key role. In that way, there also exists a link between the differential metabolic interaction network and differences in the transcriptomic or proteomic profiles of the different phenotypes.

On this aspect, kinetic models can be constructed to improve the systemic insight into a metabolic network. Over the last two decades, extensive biological studies have developed manually curated or optimized models and offered open-source files in databases such as the bio-model database (Malik-Sheriff, et al., 2020). However, a comprehensive biological modeling will need time-series experimental data; the modeling approach will be largely limited with only steady-state data. On the other hand, for large-scale kinetic models, model constructing and parameter estimating are challenging tasks (Lamichhane, et al., 2018).

Assuming that network dynamics are near steady state, biochemical interactions are represented by the system's Jacobian matrix $\boldsymbol{J}$ evaluated in steady state. Starting from metabolic networks, this paper presents a new mathematical method for analyzing the variation in the steady state Jacobian matrix of a biological system (e.g. metabolic network) between two conditions, e.g. health condition 'h' and disease condition 'd'. The method is based on a Jacobian reconstruction from the Lyapunov equation (Sun and Weckwerth, 2012)

$$\boldsymbol{J}C + C\boldsymbol{J}^T = -2\boldsymbol{D},$$

(3)

requiring only data for the computation of the covariance matrix $C$ and the estimation of a fluctuation matrix $D$.

In the last years, several works have developed inverse differential Jacobian algorithms, which provide a convenient way to infer the dynamics of metabolic networks from metabolomics data (Kügler and Yang, 2014; Nägele, et al., 2014; Steuer, et al., 2003; Sun, et al., 2015; Sun and Weckwerth, 2012; Weckwerth, et al., 2004; Wilson, et al., 2020). This method is still at an early stage and requires further improvements and unification of methods (Weckwerth, 2019). This work develops a new inverse differential Jacobian algorithm and a more extensive, automated workflow termed "COVRECON". This method combines the covariance matrix of metabolomics data with automatic metabolic interaction network modeling drawing from genome-scale metabolic reconstructions and biochemical reaction databases. As shown in figure 1b, COVRECON consists of three sub-modules: building of an organism-specific database, construction of a superpathway based metabolic interaction model (Sim-Network), and the inverse Jacobian computation.

1a,

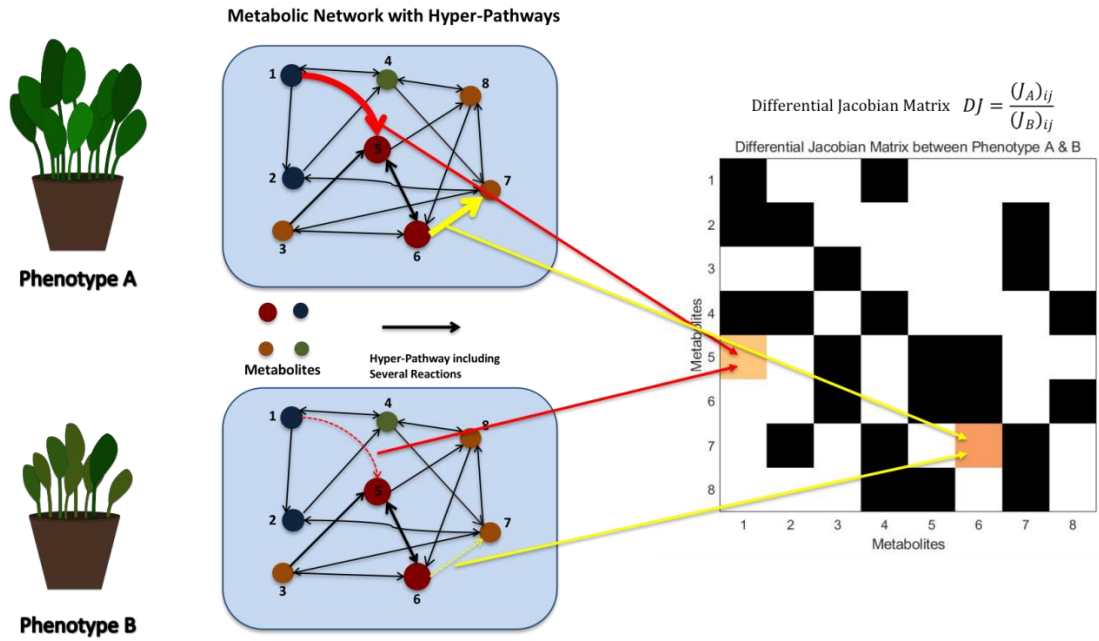

1b,

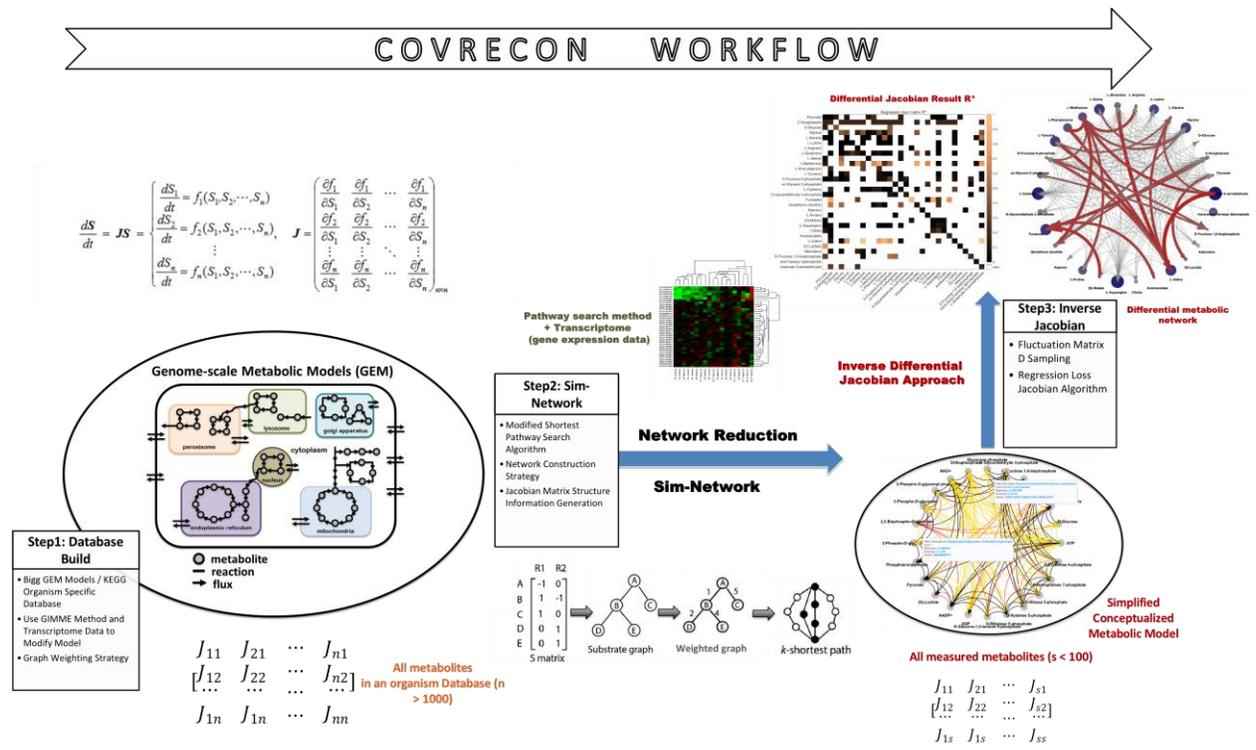

Figure1. a, Various phenotypes of an organism and differences in cellular physiology can be understood as the result from metabolic interaction network dynamics difference. This can be further uncovered with the differential Jacobian matrix. b, Work scheme of COVRECON: the covariance (COV) based differential Jacobian calculation implementing an automated metabolic network reconstruction (RECON).

## 2 Methods

### 2.1 The Differential Jacobian

Consider a biological system that consists of n compounds (metabolites, proteins) denoted by $\{X_i\}_{i=1...n}$. The system dynamics can be modeled with the set of ordinary differential equations (ODEs):

$$\frac{d\boldsymbol{M}}{dt} = \boldsymbol{F}(\boldsymbol{M}) \rightarrow \begin{cases} \frac{dM_1}{dt} = f_1(M_1, M_2, \dots, M_n) \\ \frac{dM_2}{dt} = f_2(M_1, M_2, \dots, M_n) \\ \vdots \\ \frac{dM_n}{dt} = f_n(M_1, M_2, \dots, M_n), \end{cases}$$

(1)

where $\boldsymbol{M} = \{M_i\} = \{|X_i|\}$ are the concentrations of the n compounds, and $\boldsymbol{F} = f_i(M_i)$ are composed of the reaction rates for these compounds (e.g., Michaelis-Menten kinetics, or mass action).

The steady-state Jacobian matrix $\boldsymbol{J}$ of the model is defined as a $R^{n \times n}$ matrix in which $J_{ij}$ is the first-order derivative of the rate $f_i$ for the concentration of substances $M_j$ at steady state, noted as $J_{ij} = \left.\frac{\partial f_i}{\partial M_j}\right|_{steady}$:

$$\boldsymbol{J} = \left.\frac{\partial \boldsymbol{F}}{\partial \boldsymbol{M}}\right|_{steady} = \begin{bmatrix} \frac{\partial f_1}{\partial M_1} & \frac{\partial f_1}{\partial M_2} & \cdots & \frac{\partial f_1}{\partial M_n} \\ \frac{\partial f_2}{\partial M_1} & \frac{\partial f_2}{\partial M_2} & \cdots & \frac{\partial f_2}{\partial M_n} \\ \vdots & & \ddots & \vdots \\ \frac{\partial f_n}{\partial M_1} & \frac{\partial f_n}{\partial M_2} & \cdots & \frac{\partial f_n}{\partial M_n} \end{bmatrix}_{steady}$$

(2)

Even if only evaluated at steady state, the Jacobian matrix of a system contains useful information about its dynamics, such as regulatory interactions between the different compounds. In a previous study, Steuer et al. (Steuer, et al., 2003) established a relation between the covariance matrix $C$ of the metabolic data in the network and the steady-state Jacobian matrix of the system $J$ as the Lyapunov Equation given by

$$J * C + C * J^T = -2D.$$

(3)

Thereby, $C \in R^{n \times n}$ is the covariance matrix of the compounds' concentrations $M_j$ near its steady-state value $M_j^{steady}$, and the fluctuation matrix $D$ is the covariance of noise sources acting on the system.

Here, we focus on the differences in Jacobian matrices for two biological conditions, for example a health and disease condition, abbreviated as 'h' and 'd'. Using steady-state metabolomics data of a biological network, the objective is to evaluate the differences in the Jacobian matrices and thus the changes in biochemical interactions between the two conditions.

The differences between the two conditions are quantified by the differential Jacobian $DJ$, the elements of which are defined from the Jacobians in first, e.g., health, condition 'h' $J_h$ and in the second, e.g., disease, condition 'd' $J_d$ as

$$DJ_{ij} = \begin{cases} \left|\frac{(J_d)_{ij}}{(J_h)_{ij}}\right| \\ 1, \quad if\ (J_h)_{ij} = 0. \end{cases}$$

(4)

In the Lyapunov Equation Eq. (3), the Jacobian matrix $J$ has $n \cdot n$ unknown variables, while the covariance matrix $C$ is a symmetric matrix, and thus has only $\frac{n(n+1)}{2}$ independent variables to be determined from measurement data. This fact indicates that in the direct inverse approach of calculating $J$ from $C$, we would generally have only $\frac{n(n+1)}{2}$ equations but $n \cdot n$ unknown variables, thus the inverse Jacobian approach would generally be under-determined and the reconstructed Jacobian matrix non-unique.

However, for realistic biological networks the Jacobian matrix structure is commonly a sparse matrix (Nägele, et al., 2014; Sun and Weckwerth, 2012). Thus, the first step of COVRECON, Sim-Network, is made to automatically construct a metabolic interaction network model for the measured metabolites, which is used to constrain the Jacobian matrix structure. In most cases, the resulting structure leaves fewer non-zero entries than independent variables in the covariance matrix, thus making the Lyapunov equation over-determined and requiring the use of regression methods.

The COVRECON method consists of two major steps:
1. Determine the Jacobian structure. The key objective is to reduce the network structure from a genome-scale network to the specific biochemical species that are included in the considered dataset.
2. Analyze the differential Jacobian matrix. We establish a new method that focusses on identifying the major components in the differential Jacobian, instead of pursuing a full quantitative reconstruction.

The following sections will describe these two steps in more detail.

## 2.2 Determination of the Jacobian structure with the software Sim-Network

The Jacobian matrix structure is determined by the software tool Sim-Network which we implemented in Matlab. It constructs a conceptualized metabolic model for the metabolites in the considered dataset using a pathway search approach. Methods for pathway design and prediction have a long history since the emerging genomics, proteomics, and metabolomics databases (e.g., BIGG models (King, et al., 2016), KEGG (Ogata, et al., 1999), Metacyc (Caspi, et al., 2020), and ModelSEED (Seaver, et al., 2021)).

To construct the conceptualized metabolic model, Sim-Network makes use of reaction data from BIGG genome-scale models (King, et al., 2016), the KEGG database (Ogata, et al., 1999). For relevant reactions thermodynamics data (reaction irreversibility, direction and estimated reaction delta Gibbs free energy), we utilize the ModelSEED dataset (Seaver, et al., 2021).

Figure 1b illustrates a scheme of the Sim-Network tool; the aim of Sim-Network is reducing the genome-scale model (with more than 1000 metabolites) to a superpathway based metabolic interaction network for the measured metabolites set (with less than 100 metabolites). In general, Sim-Network contains three main steps: network information gathering, path search and pruning. We first generate a reaction database from a BIGG model or all the reactions related to a specific organism in KEGG. Then Sim-Network will gather relevant metabolites, and assemble a directional weighted network representation. Next, through shortest path search algorithm, Sim-Network will compute shortest paths (in both directions) for all pairs of metabolites in a specific metabolomics dataset. Finally, regarding the predefined cost threshold, Sim-Network will prune the network (assuming long-distance interactions are negligible), and construct a metabolic interaction network for the considered dataset. The detailed workflow of Sim-Network consists of the following steps:

Step 1: Initially, one needs to choose an organism-based genome-scale metabolic model as a database. If transcriptomic data is available, the model is modified and trimmed by discarding reactions whose enzyme is not activated, using the GIMME method (Becker and Palsson, 2008). When no transcriptomic data is available, the KEGG database offers a broader database including more enzyme specific reactions. By choosing the organism in COVRECON toolbox according to the experimental dataset (e.g. hsa for Homo sapiens, mmu for Mus musculus), Sim-Network will search and include all the enzymes and reactions of that organism into a database model.

Step 2: The model stoichiometric matrix is transformed into its network representation. Meanwhile, a weighting strategy is applied to the reactions. Firstly, the side-metabolites (e.g., H2o, H+...) are excluded during the pathway search. This is because those side-metabolites concentrations are influenced by many other metabolites, which makes the influence from each specific metabolite negligible. The predefined side-metabolites are listed in the supplemental material. For each reaction, the reaction direction and irreversibility information is obtained from the Bigg model or from the ModelSEED dataset. The forward direction is given weight 1. If the reaction is reversible, the reverse reaction is given a user-defined weight (default 2). If the Gibbs free energy difference of the reaction $\Delta G$ is available in ModelSEED database, we adjust the path cost for the reverse reaction according to the strategy in MRE web tool (Kuwahara, et al., 2016). Generally speaking, for the reverse reaction whose Gibbs free energy difference is larger than 100 kcal/mol, we will add the log value of the delta Gibbs free energy to the user-defined reverse reaction weight. Based on the weighting strategy, Sim-network will connect every pair of reactant-product in each forward or reverse reaction with directional connections, and give the related reaction weight to the directional connections. Since the reaction rate is influenced by all reactants, two metabolites are also connected in Sim-Network when they are both reactants in a reaction, or both products in a reversible reaction.

Step 3: For the selected metabolite subset $\{X_s\}$ of the database metabolites set $\{X\}$, we determine and save all the shortest path costs and routes for every directional pair of metabolites in $\{X_s\}$. Sim-Network discards all routes of which the cost is higher than a preset threshold (default is 3). Then based on the search result, we construct the metabolic interaction model and generate the related Jacobian matrix structure based on a simplified model strategy. For example, if the shortest route from metabolite $X_a$ to $X_c$ contains another metabolite $X_b$ in the selected metabolites set $\{X_s\}$, we will discard this connection, assuming that the influence from $X_a$ to $X_c$ can be reflected by the co-influences of ($X_a$ to $X_b$) and ($X_b$ to $X_c$). The resulted network is saved in SBML format (Hucka, et al., 2003) with the Sim-Biology Matlab toolbox.

With the conceptualized network, one can introduce the Jacobian structure. The diagonal components of the related Jacobian matrix are non-zero components, and each connection in the conceptualized network represents a non-zero off-diagonal component. In Supplemental material S3, we provide a toolbox manual and case studies.

**2.3 The regression loss matrix as a new inverse differential Jacobian evaluation**

This section introduces a new algorithm to determine the major components of the differential Jacobian matrix, using the Jacobian structure information as determined by Sim-Network and a covariance estimation from metabolomics data.

In the inverse Jacobian approach, the Lyapunov equation (3) is solved for the Jacobian matrix $J$ with given covariance and fluctuation matrices $C$ and $D$. This is a linear equation, which as outlined in previous studies (Kügler and Yang, 2014; Sun and Weckwerth, 2012) can be rewritten

in the form

$$A q = b,$$

(7)

where the vector $q$ consists of all the non-zero components in the Jacobian matrix as unknown variables. The matrix $A$ is calculated from the values in the covariance matrix $C$; its dimension is $(\frac{n(n+1)}{2}, L)$, where $L$ is the dimension of the vector $q$. Since the Jacobian structure is sparse, we usually have $L < \frac{n(n+1)}{2}$, and equation (7) is overdetermined. The vector b is constructed from the fluctuation matrix $D$; its dimension is $\frac{n(n+1)}{2}$.

Similarly, to evaluate the differential Jacobian in two conditions denoted by 'h' and 'd', the corresponding Lyapunov equations can be rewritten as

$$A_h \, q_h = b_h$$
$$A_d \, q_d = b_d.$$

(8)

Previous inverse Jacobian algorithms (Kügler and Yang, 2014; Nägele, et al., 2014; Steuer, et al., 2003; Sun, et al., 2015; Sun and Weckwerth, 2012; Weckwerth, et al., 2004; Wilson, et al., 2020) have assumed that independent stochastic noise affects each metabolite individually, giving rise to a diagonal fluctuation matrix $D$ in the Lyapunov equation (3). Thus these studies used an average result of randomly sampled diagonal perturbation matrices $D$ (Nägele, et al., 2014; Sun and Weckwerth, 2012), or applied a L-p optimization of the two conditional fluctuation matrixes $D_h$ and $D_d$ (Kügler and Yang, 2014) to calculate the differential Jacobian matrix $DJ$. These methods work through directly solving the linear equation (7). However, when the condition number of the matrix $A$ is large, the linear equations' solution will be unstable against small perturbations in the data. This makes previous methods (Kügler and Yang, 2014; Nägele, et al., 2014; Sun, et al., 2015; Sun and Weckwerth, 2012) numerically sensitive and not adequate to deal with large-scale models.

As an alternative to a direct solution of the linear equations (8) and subsequent calculation of the differential Jacobian according to (4), we introduce the use of the regression loss as a measure for the relevance of individual components in the differential Jacobian. Solving an overdetermined linear equation of the general form as in (Freedman, 2009) by linear regression, we obtain the solution

$$q^* = (A^T A)^{-1} A^T b,$$

and we define the (linear) regression loss $r$ as

$$r = \|b - Aq^*\|.$$

As shown in the supplementary material, under numerical variations in $b$ the variation of the regression solution $q$ is proportional to the condition number of $A$, while the variation in the regression loss $r$ does not scale with the condition number.

This property can be used to make the determination of large elements in the differential Jacobian more robust against fluctuations in the data. To this end, we construct a *regression loss matrix R* that aims to capture the relative importance of individual elements in the differential Jacobian. The regression loss matrix has the same dimension and sparsity structure as the Jacobian $J$, determined by Sim-Network. Each non-zero element of the regression loss matrix is computed as in Equation (9), where $A_c$ is calculated by combing $A_h$ and $A_d$ in Equation (8) with the additional constraint that only that single element $J_{ij}$ may differ between the Jacobians $J_h$ and $J_d$, while all other elements are equal.

$$q_s^* = (A_c^T A_c)^{-1} A_c^T b_s$$
$$R_{ij}^* = \min_{b_s} \|b_s - A q_s^*\|$$

*(9)*

To solve the regression problem, one needs to use specific values for the fluctuation matrices $D_1$ and $D_2$. However, in practice these are not known. Therefore, similar as in previous studies (Nägele, et al., 2014; Sun and Weckwerth, 2012), we sample over possible values $b_s$ of the fluctuation matrices (diagonal elements distributed between 0 and 1), and take the minimum regression loss as the overall result for the regression loss matrix as in Equation (9). A more comprehensive description of the new algorithm is presented in Supplemental Material S1.

For comparison, we replicated the L-p optimization used in the previous work [19]. The general idea of the L-p optimization is that the L-p cost optimization will induce most components in the optimized differential Jacobian matrix (refer to (DJ-1) in equation 4) close to zero.

Furthermore, previous studies do L-p optimization based on diagonal D matrix; we extended it into diagonal-dominant matrix optimization and applied an improved algorithm compared to [19], which integrates several global optimization approach. The details of the original and improved L-p optimization approach are presented in Supplemental Material.

**2.4 The inverse differential metabolic interaction network**

The differential metabolic interaction network illustrates the change of the metabolic network between two phenotypes and are visualized by circular interaction plots. Here, each node i represents a metabolite in the dataset; the thickness of the line $j \rightarrow i$ is proportional to $DJ_{ij}$ in Eq. (4) or entry $R^*_{ij}$ of the regression loss matrix (9); the node size $i$ is proportional to $DJ_{ii}$ in Eq. (4) or the entry $R^*_{ii}$ of the regression loss matrix (9). The generation of these circular plot is included in the COVRECON toolbox with different setting for the resulting differential metabolic interaction network as shown in the Supplemental material S3 (toolbox manual). The detailed enzyme and gene information of each superpathway can be checked interactively in the circular plot.

**2.5 Model case studies**

To evaluate the new regression loss Jacobian algorithm, we utilize an abstract test model and several published models obtained from the EBI BioModels database (Malik-Sheriff, et al., 2020). The following models are utilized in this evaluation, using reaction perturbations as described to obtain the two conditions ('h' and 'd'):

1. Abstract test model with six components (details see supplemental material).

2. Model of the upper glycolysis pathway from Klipp et al. (Klipp, et al., 2016): Similar to previous work in Kugler et al. (Kügler and Yang, 2014), in order to mimic a second network condition, we introduced a twofold increase of the phosphorylation rate parameter k4 of reaction $R4: Fruc6P + ATP \rightarrow Fruc1,6P_2 + ADP$ from its nominal value $k_4^h = 1$ to $k_4^d = 2$. Figure 2a shows the exact differential Jacobian matrix $D\mathbf{J}$ between these two conditions at the steady state.

3. Model of the EGFR/ERK signaling pathway in Orton et al. (Orton, et al., 2009): This work studied the mutation of SOS feedback reactions. The differential Jacobian matrix of the wild type and mutation models is shown in Figure 3a. These two models are also used as evaluation models in (Kügler and Yang, 2014).

4. Mathematical model of carbohydrate energy metabolism (Nazaret and Mazat, 2008): We increased the reaction rate parameter in $R2: Pyrute + NADHc \rightarrow Lactose + NADc$ five-fold for the second conditional Jacobian matrix. Figure 2c shows the exact differential Jacobian matrix for this model.

5. AMPK-mTOR pathway model (Dalle Pezze, et al., 2016): The paper describes a wild type and mTOR knockout model based on time-series experimental data, which represent our two conditional Jacobian models. The exact differential Jacobian matrix is in Figure 3b.

6. Hepatic glucose metabolism model (Bulik, et al., 2016). We applied a two-fold parameter change for the reaction rate of the second reaction R2 to generate the second conditional model. The related differential Jacobian matrix is shown in Figure 3c.

7. Large-scale blood cell metabolism model (Holzhütter, 2004). We introduced a five-fold increase to several components of the Jacobian matrix directly; the exact differential Jacobian matrix is shown in Figure 3d.

Table 1. Overviews of the evaluation models.

| Model type | Independent components number | Reaction number | Non-zero Jacobian components number | Matrix condition number scale |
|---|---|---|---|---|
| Test model | 6 | 7 | 20 | 1.00E+02 |
| The upper glycolysis pathway model | 5 | 8 | 16 | 5.00E+01 |
| EGFR/ERK signaling pathway model | 13 | 13 | 31 | 1.00E+04 |
| Carbohydrate energy metabolism model | 12 | 18 | 59 | 1.00E+06 |
| AMPK-mTOR pathway model | 31 | 48 | 159 | 1.00E+05 |
| Hepatic glucose metabolism model | 27 | 35 | 108 | 1.00E+10 |
| Blood cell metabolism model | 33 | 29 | 252 | 1.00E+10 |

# 3 Results

## 3.1 The correlation matrix does not recover the Jacobian

Correlation network topology analysis is one of the frequently used multivariate statistical methods to study metabolic interactions (Kitagawa, et al., 2019; Poldrack, et al., 2015; Tofte, et al., 2019; Weckwerth, 2010; Weckwerth, 2011; Weckwerth, et al., 2004). However, correlation network analysis alone is not suitable to study the differential metabolic regulation between two conditions. Using the models 2, 4, 6 and 7, we generated the differential correlation matrixes under the two conditions. Then, we compared the differential Jacobian matrices and the differential correlation matrices (supplementary Figure 1). The results show that the differential correlation matrix cannot recover the differential Jacobian matrix. Here we note that, a correlation matrix or network can still be utilized to show potential interactions when the Jacobian structure information is not clear (Kitagawa, et al., 2019; Poldrack, et al., 2015; Tofte, et al., 2019).

## 3.2 The regression loss Jacobian is more reliable than the L-p based Jacobian

In the following, we describe the results of applying either the regression loss Jacobian algorithm or the improved L-p optimization algorithm for the differential Jacobian reconstruction on the different models as described in Section 2.4. Since in previous studies (Kügler and Yang, 2014; Nägele, et al., 2014; Steuer, et al., 2003; Sun, et al., 2015; Sun and Weckwerth, 2012; Weckwerth, et al., 2004; Wilson, et al., 2020), the fluctuation matrix $D$ is assumed to be a diagonal matrix, we restrict this comparison to a diagonal structure of $D$.

In this evaluation, artificial data is generated using Gaussian perturbations acting on individual compounds only for all the evaluation models in Section 2.3.3. The covariance matrices for models 1 and 2 are computed through SDE simulation with 100 and 300 samples, respectively, while in other models the covariance matrices are determined from the Lyapunov equation with $\varepsilon_D = 0.4$ (Section 2.4). The Jacobian reconstruction made use of structure information derived directly from the underlying model. Both the L-p optimization method and the new regression loss Jacobian algorithm were applied to that data. As shown in supplementary Table 1, the new inverse method achieves a more accurate result with less computation time than the L-p optimization. The table shows the accuracy of three inverse differential Jacobian approaches: Kugler et al. (Kügler and Yang, 2014) (details in Supplementary Material), an improved L-p optimization approach and our new inverse Jacobian algorithm. The indicated cost is the L-p optimization loss; the target cost is the L-p loss calculated with the real $b_h$ and $b_d$ (detail in Supplemental Material). From the table, we can see that the improved L-p optimization approach can obtain a better result than Kügler et al. (Kügler and Yang, 2014), but requires more computation time. Our new inverse Jacobian algorithm will need fewer samples to reach a similar accuracy (100 compared to 1000) while also cutting down the computation time.

The inverse differential metabolic interaction network for both the L-p algorithm and our new regression loss Jacobian algorithm with all the models are shown in Figure 2. The results show that because of the instability of the regression solution with ill-conditioned matrices, the L-p optimization will be inadequate when either the condition number $\boldsymbol{K}$ $(J)$ of the Jacobian is large, or the model dimension is large. The regression loss Jacobian algorithm can achieve a better accuracy and numerical stability by utilizing the regression loss matrix $R^*$ to recover the relevant components of the differential Jacobian $D\boldsymbol{J}$. The actual matrices (in place of the circular interaction plots) are shown in supplementary Figure 2. Furthermore, supplementary Figure 3 shows a scatter plot between the real differential Jacobian components values, scaled to the interval (0, 1) and the calculated $r(i,j)$ in our regression loss Jacobian algorithm for all models. This confirms that with enough samples, the regression loss Jacobian algorithm can find most differential Jacobian components for the two conditional Jacobian matrixes, except for some false negative components in the larger models 6 and 7.

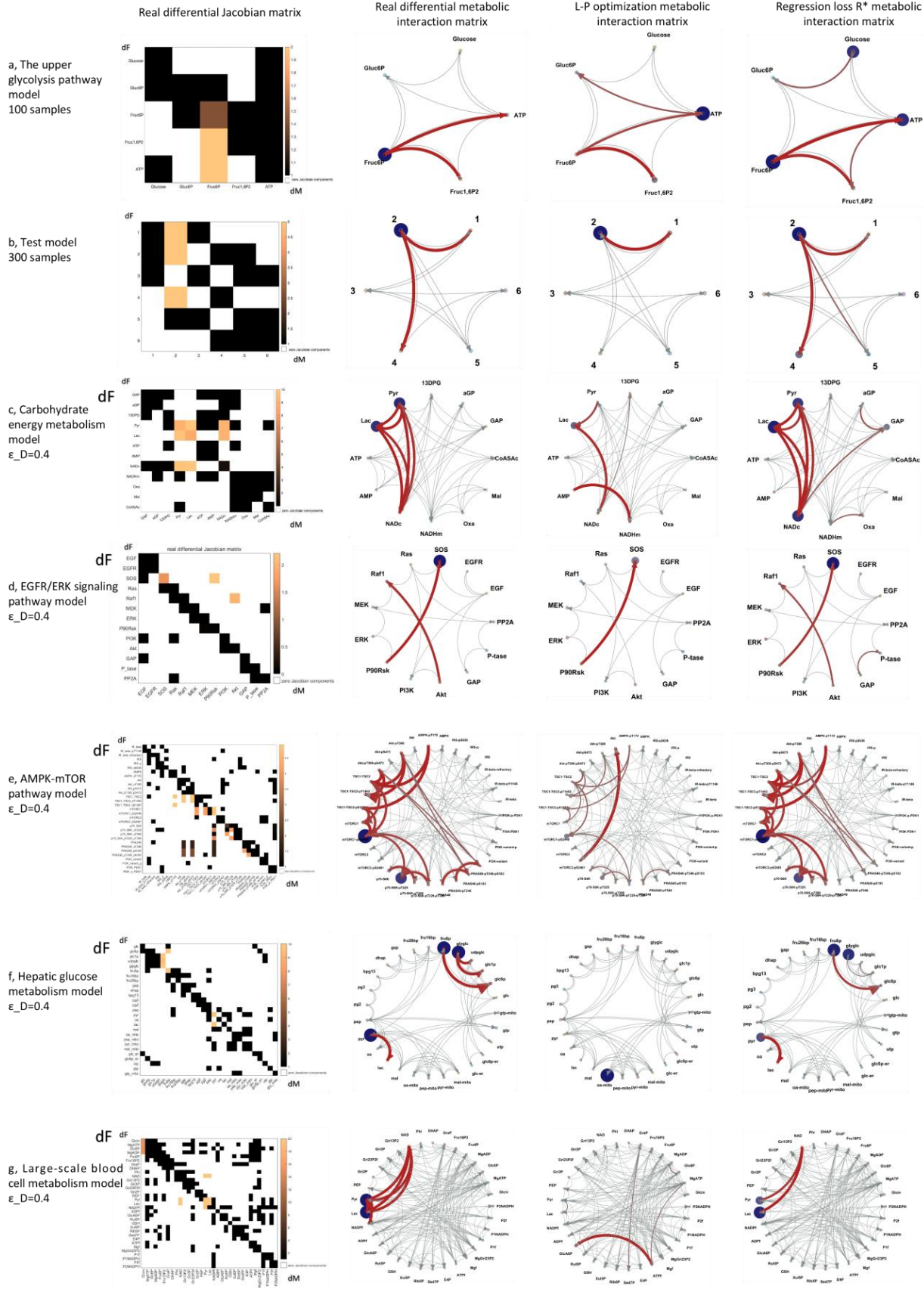

Figure 2. The inverse differential metabolic interaction network with both L-P optimization and the new regression loss Jacobian algorithm for all evaluation models in section 2.5. The evaluation models are: a, textbook model of the upper glycolysis pathway; b, test model; c, carbohydrate energy metabolism model; d, EGFR/ERK signaling pathway; e, AMPK-mTOR pathway model; f, hepatic glucose metabolism model; g, large-scale blood cell metabolism model. In each sub-graph, the left two subplots gives the real differential Jacobian matrix and differential interaction network; the right two subplots show the calculated differential interaction network through the L-p optimization approach and the regression loss Jacobian algorithm respectively (refer to Section 2.3).

To test the numerical stability of the new algorithm, we calculated the statistical accuracy by using different noise levels in the ***D*** matrix, $\varepsilon_D = 0.2, 0.3, 0.4$ and $0.5$. For each model and $\varepsilon_D$, we repeated the evaluation 100 times and calculated the replicability of the top 1, 3 and 5 differential Jacobian components in $R^*$ (refer to Eq. (S4*)). The results are listed in Table 2, where we observe that the new algorithm can identify the top 1, 3, and 5 components with high accuracy. Of note, even in the larger models 6 and 7, which have several false negatives, these false negatives still have a high consistency, indicating that the noise level in the ***D*** matrix does not significantly impact the results.

Table 2. The regression loss Jacobian 100 repeats statistics results with diagonal theoretical D for the evaluation models 4-7 as in section 2.3.3.

| Evaluation model | Accuracy in 100 repeats (top1, top3, top5) | | | |
| --- | --- | --- | --- | --- |
| | $\varepsilon\_D=0.2$ | $\varepsilon\_D=0.3$ | $\varepsilon\_D=0.4$ | $\varepsilon\_D=0.5$ |
| Carbohydrate energy metabolism model | 1, 0.99, 0.96 | 0.88, 0.88, 0.84 | 0.86, 0.83, 0.76 | 0.72, 0.70, 0.63 |
| AMPK-mTOR pathway model | 1, 1, 1 | 1, 1, 1 | 1, 1, 1 | 1, 1, 0.99 |
| Hepatic glucose metabolism model | 1, 1, 0.98 | 1, 1, 0.96 | 0.99, 0.88, 0.84 | 0.97, 0.81, 0.84 |
| Blood cell metabolism model | 1, 1, 0.95 | 1, 1, 0.93 | 1, 0.96, 0.88 | 0.97, 0.95, 0.84 |

Finally, we tested the regression loss Jacobian method with covariance data generated from stochastic simulations for the models 1, 3, 4 and 5 with a stochastic second-third order Runge-Kutta implicit method. For each model, we added the stochastic Gaussian noise perturbations to every component at each time step (Higham, 2008), which corresponds to the nominal ***D*** matrix being a diagonal matrix. For each model, we repeated the computation with covariances computed from 100 and 1000 samples to evaluate the effect of sample size on the results. Supplementary Figure 4 demonstrates the results of the new inverse Jacobian algorithm. As shown in Supplementary Figure 4a and Figure 4b, for small models (about 10 variables) the algorithm correctly finds the large differential Jacobian components through large values in the regression loss $R^*$ using only 100 samples. For larger models, the algorithm is only able to find some of the relevant differential Jacobian components with 100 or 1000 samples, as shown in Supplementary Figure 4c and 4d. Even though the detection is improved with 1000 compared to 100 samples, there are still some false negatives even with 1000 samples. We can conclude that this new inverse Jacobian algorithm gives reliable results using on the order of 100 samples for metabolic models with about 10 variables. For larger models, we expect the required sample number to grow with the square of the model size, presenting a practical challenge for larger models. However, even with an insufficient number of samples, we can still find several relevant differential Jacobian components through large values of the regression loss $R^*$.

### 3.3 COVRECON workflow case study

As in the first step of COVRECON, Sim-Network will reconstruct a reduced model; we need to test the influence of this network reduction approach (see "5 Effect of the network reduction on the Jacobian reconstruction" in the Supplemental material S1. The result verifies the feasibility of this first step (Sim-network) in the COVRECON approach. Even when using the reduced network structure for the Jacobian reconstruction, the algorithm is able to detect most of the relevant interactions in the original network.

In the next step, the complete COVRECON workflow is tested starting from a data covariance matrix, without additional structure information. The test is done with models 6 and 7 (hepatic glucose metabolism model the blood cell metabolism model). With these models, we generate the covariance matrices $C_h$ and $C_d$ for both conditions 'd' and 'h' with the Lyapunov equation using $\varepsilon_D = 0.5$. In the reconstruction, we first use Sim-Network to determine the Jacobian structure information, relying biochemical pathway information in the KEGG database for *Homo sapiens* (HSA). As algorithm parameters, we use a cost threshold of 2; discard side-metabolites (listed in supplementary material); apply reverse reaction weight of 2; and use thermodynamics to modify the reverse reaction weight. The reconstructed metabolic interaction networks and Jacobian structure matrices of the two models are shown in Figure 3.

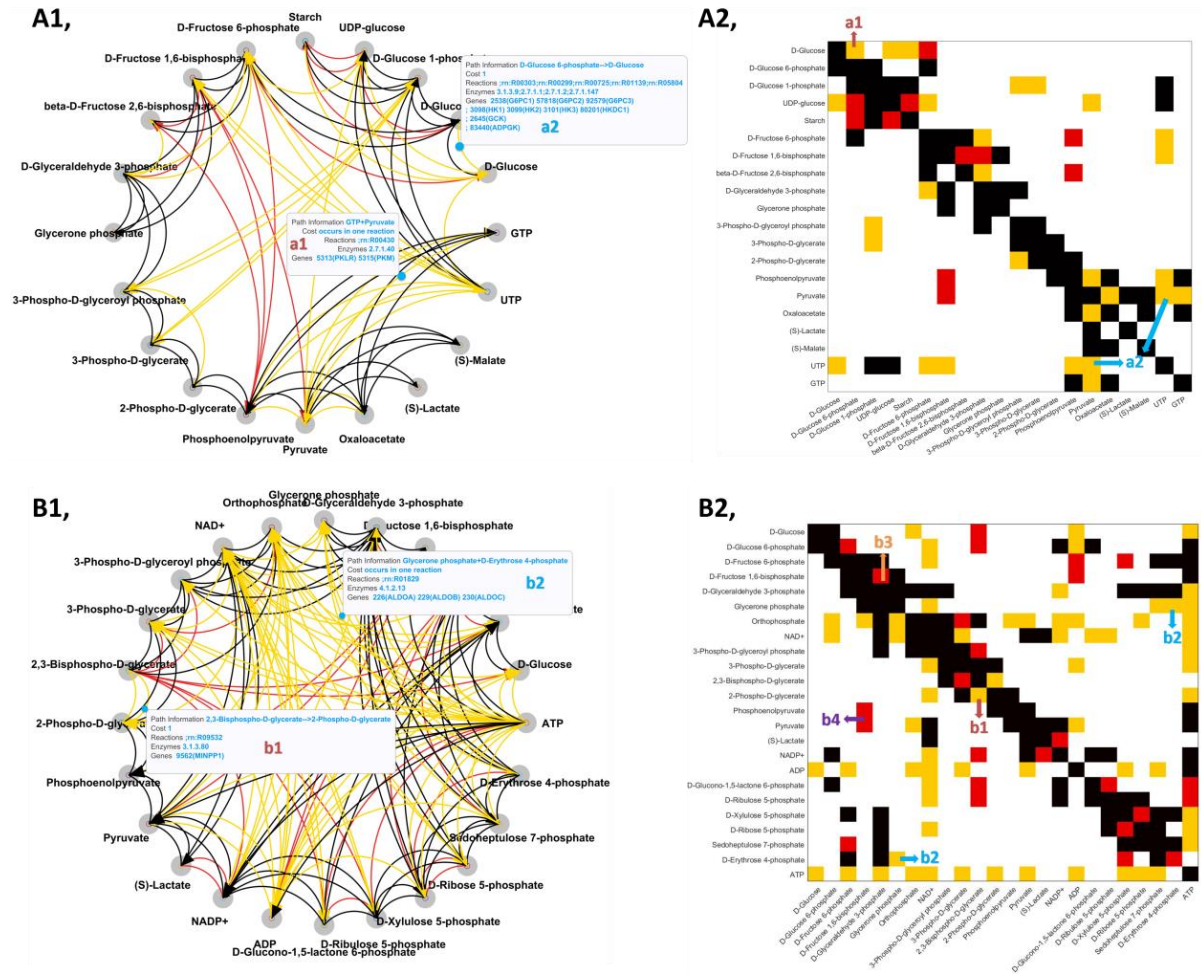

Figure 3. Metabolic interaction network reconstruction and Jacobian structure matrix with Sim-network tool. A and B give two example cases for the case study of Sim-network. A1 and B1 present the reconstructed networks. A2 and B2 show the related Jacobian structure information compared to the literature Jacobian structures. An orange Jacobian line (left subplots) or component (right subplots) represents a new added non-zero Jacobian interaction, a red Jacobian line (left subplots) or component (right subplots) represents an unconnected non-zero Jacobian interaction. The marked Jacobian components a1, a2 and b1-b4 are the typical cases for illustration.

Generally we find that the network structure determined by Sim-network contains more interactions than the related literature model. A few examples are labelled in Figure 3 and discussed in the following. Labels a1 and b1 correspond to one-step reaction interactions: a1, D-

Glucose 6-phosphate to D-Glucose (KEGG R00303) and b1, 2,3-Bisphospo-D-glycerate to 2-phospo-D-glycerate (KEGG R09532). In addition, in cases a2 and b2, the metabolites are connected because they occur together in a reaction: a2, Pyruvate and UTP in (KEGG R00659) and b2, Glycerone phosphate and D-Erythrose 4-phosphate in (KEGG R01829). These reactions are not included in the literature models, but are registered in KEGG and are thus found by applying Sim-Network. On the other hand, in case b3, the connection from D-Glyceraldehyde 3-phosphate (GraP) to D-Fructose 1,6-bisphoaphate (Fru16P2) is found by Sim-Network as GraP->D-Fructose 6-phosphate (Fru6P)-> Fru16P2 (step1: KEGG R08575, step 2: KEGG R00756); but it is subsequently discarded as an indirect interaction since the intermediate metabolite Fru6P is also in our network (according to the network simplification strategy explained in the method part). As for case b4, the Jacobian interaction between Fru16P2 and Pyruvate & Phosphoenolpyruvate (PEP) is not discovered by Sim-network. This is because Fru16P2 will activate the catalyzing enzyme Pyruvate kinase through an allosteric regulation, but Sim-network does currently not take interactions resulting from enzyme regulations into account. The circular plot in matlab format is in Supplemental material S4 & S5, where the detailed information of all superpathways can be checked. In addition, the text version of the superpathways information is presented in Supplemental material S2.

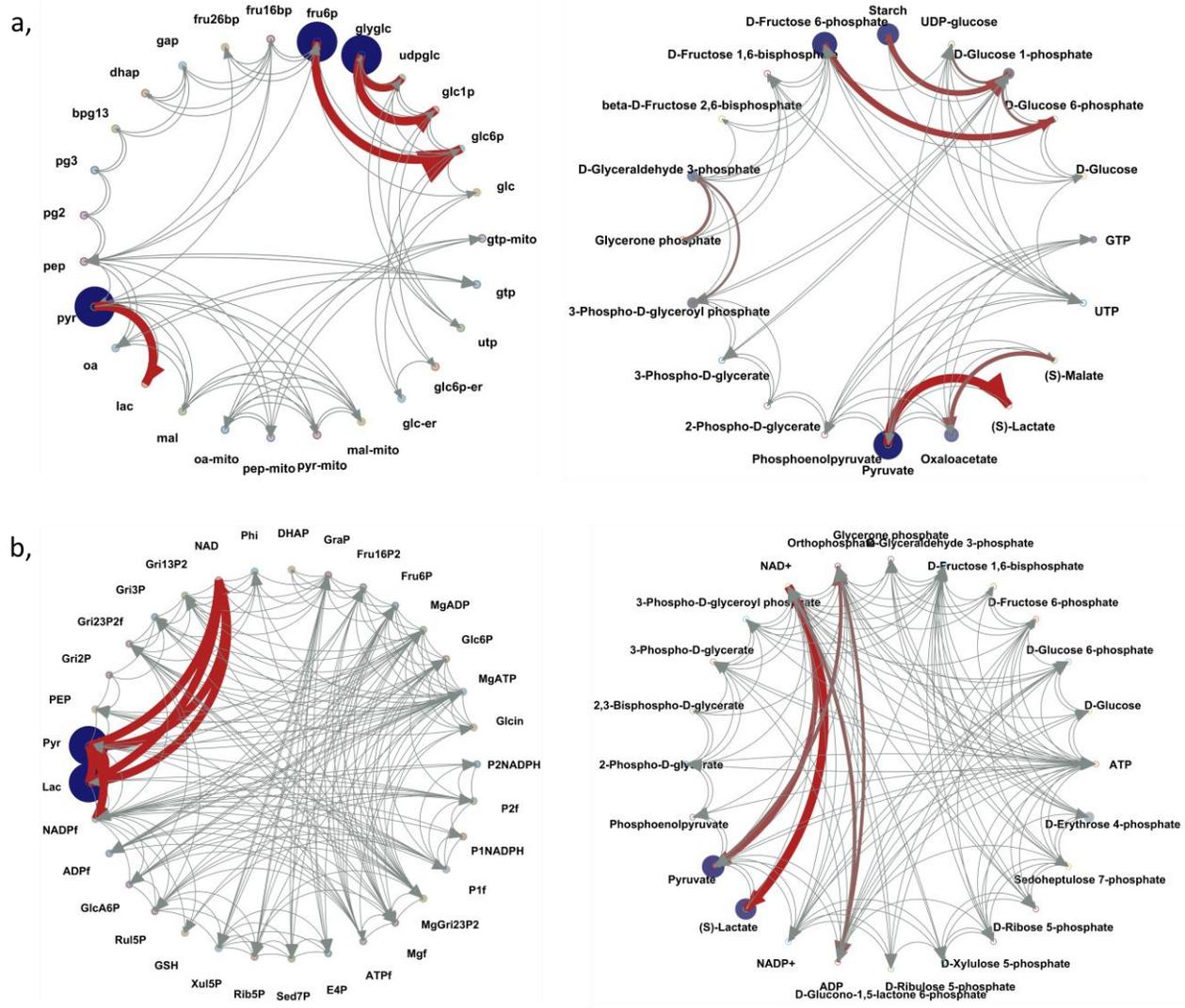

Figure 4. Complete workflow of the COVRECON strategy. SIM-Network is used for network reduction and metabolic interaction matrix generation. Subsequently, a covariance data matrix is used to calculate the differential Jacobian in form of the Regression loss matrix $R^*$. Inverse differential metabolic interaction network determined from COVRECON (right) and the exact differential metabolic interaction network (left). The circular plot in matlab figure format is in Supplemental material S6 & S7.

In conclusion, compared with manually built models, Sim-Network has several advantages. First, the resulting model is all-inclusive, for it will explore all the reactions in one specific organism and locate all possible routes, without relying on a modeler's domain knowledge. On the other hand, the automatically constructed model will be more complex than a manually built model, and it might include rare reactions and routes. Domain knowledge could thus still be helpful to remove individual interactions, starting from the detailed information given by the Sim-network reconstruction.

Moreover, the enzyme regulatory interactions are not included in the current tool. In fact, these interactions are still rarely known except for several widely studied enzyme activators or inhibitors such as fructose bisphosphate for pyruvate kinase. Future versions of COVRECON could make use of an enzyme regulation database to also take these interactions into account.

With the reconstructed network information, we then apply the regression loss Jacobian algorithm to the covariance matrices. As shown in figure 4 and Supplemental Figure 5, the COVRECON approach shows similar results to the results with exact Jacobian structure information obtained from the model. As a further evaluation, we also consider different Sim-network settings to reconstruct the network (cost threshold 1, reverse reaction weight 1, no thermodynamic strategy, see supplementary material "6 Sim-Network Matlab Interface and default settings"). This will result in more connections because we treat the forward and reverse reaction direction in the same way. As shown in supplementary Figure 6, with this setting, the reconstructed model will miss fewer components present in the literature model but have even more components which are not present in the literature model. The resulting regression loss matrix $R^*$ remains similar in each case. Overall, this analysis verifies that the COVRECON workflow can recover relevant interactions in a differential Jacobian from only metabolite covariance data.

### 3.4 Application to an experimental dataset from literature

Finally, we applied the COVRECON to a real experimental dataset from a breast cancer study (Di Filippo, et al., 2022). We analyzed the differential Jacobian matrix between two dataset: non-tumorigenic breast epithelial cell line (MCF102A) and pleural effusion metastasis of a breast adenocarcinoma (MCF7). For the reconstruction, we use the KEGG dataset with organism homo sapiens (KEGG code: hsa), the Sim-network settings are left at their default values, and the transcriptomic dataset is used to discard inactivate reactions with GIMME method (refer to method section 2.2). The discovered superpathways with relevant reactions, enzymes and genes information are listed in the Supplemental file S2. A COVRECON toolbox manual of the case study is presented in Supplemental file S3.

Figure 5 illustrates the inverse differential metabolic interaction network, where highlighted components imply major differences in metabolic interactions between the two datasets. Here, different from Section 2.4, the node size represents the Variable Importance (calculated as -log(p), where p is the p-value of a t-test from the metabolomics datasets comparison). The corresponding matrix is presented in Supplemental Figure 7. Moreover, we attached the detailed enzyme gene information of each metabolic interaction line in Supplemental material S8, where one can interactively check detailed information about each metabolic interaction. In addition, the information is also presented in Supplemental material S2. As shown in Figure 5, we found several metabolic interactions with a major difference between the datasets, in which also enzyme transcript levels underlying these interactions had a significantly different expression level. In Supplemental Figure 8, we list all the t-test results of the transcriptomic profile for interactions with highlighted value above 0.5.

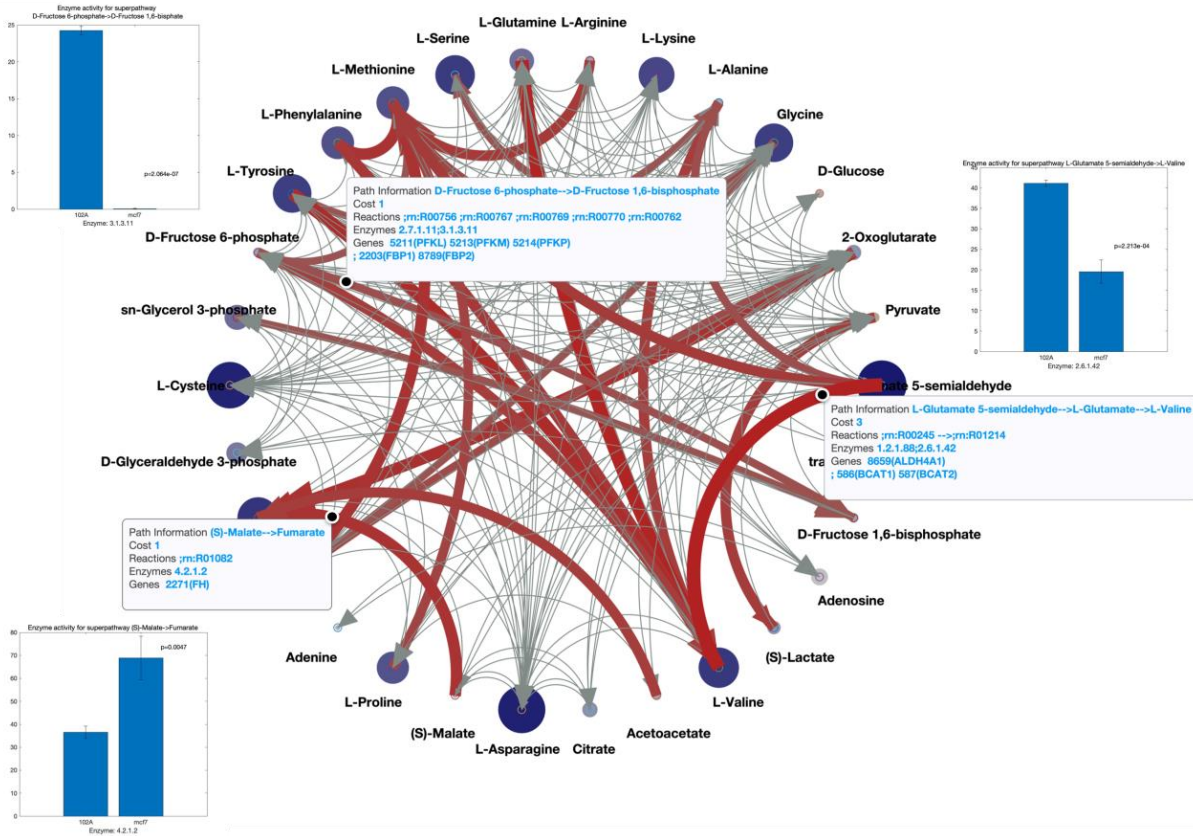

Figure 5. The differential metabolic interaction network plot, where the node size represents the Variable Importance (calculated as -log(p), p is the p-value of t-test from metabolomics datasets comparsion), and the line widths describe the metabolic interaction difference. We chose and list the superpathways details of three highlighted metabolic interactions and performed a t-test for relevant transcriptomic data. The t-test results show significant ($p<0.05$). The matlab format circular plot with all superpathway information is in Supplemental material S8.

## 4 Conclusion

In this paper, we have developed a new approach for the inverse differential Jacobian algorithm: COVRECON. Unlike the widely used constraint-based analysis for the reaction-flows optimization, this approach endeavors to discover the causal biochemical interactions between two conditions of the system. This new approach offers an alternative mathematical approach to process and interpret large-scale metabolomics data. The open-source matlab-tool is available in the website https://bitbucket.org/mosys-univie/covrecon. Supplemental file S3 gives a toolbox manual.

The main subject of this new approach is large-scale metabolomics data. Meanwhile, we also tried to integrate different OMICS datasets. In Sim-Network, the transcriptomic data can be used to exclude reactions with low activity; and the important enzyme regulations can be added into pathway search for a better network reconstruction.

From our knowledge, COVRECON is the first method to integrate different OMICS data and automatically construct a metabolic interaction model providing a more general network structures and perturbations of the same. As for the inverse Jacobian part, this work introduces a novel algorithm improving accuracy and stability, reduces computation time, and extends the method to large-scale models.